\documentstyle[aps]{revtex}
\input epsf

\begin{document}
\title{Peculiarities of the resistive transition in fractal superconducting
structures}
\author{Yuriy I. Kuzmin}
\address{Ioffe Physical Technical Institute of the Russian Academy of Sciences,\\
Polytechnicheskaya 26 St., Saint Petersburg 194021 Russia,\\
and State Electrotechnical University of Saint Petersburg,\\
Professor Popov 5 St., Saint Petersburg 197376 Russia\\
e-mail: yurk@mail.ioffe.ru; iourk@yandex.ru\\
tel.: +7 812 2479902; fax: +7 812 2471017}
\date{\today}
\maketitle

\begin{abstract}
The influence of fractal clusters of a normal phase on the current-voltage
characteristics of a percolation superconductor in the region of a resistive
transition has been studied. The clusters represent the aggregates of
columnar defects, which give rise to a correlated microscopic disorder in
the system. Dependencies of the static and dynamic resistance on the
transport current are obtained for an arbitrary fractal dimension of the
cluster boundaries. It is revealed that a mixed state of the vortex glass
type is realized in the superconducting system involved.
\end{abstract}

\bigskip

Fractal superconducting structures possess a number of unusual magnetic and
transport properties \cite{pla} -\cite{pla2}. The possibility to enhance
pinning by trapping the magnetic flux in the normal phase clusters with
fractal boundaries \cite{pla1}-\cite{pss} is of much interest. The study of
peculiarities of the current-voltage ({\it V}-{\it I}) characteristics in
the region of a resistive transition allows getting new information about
the nature of the vortex state in such systems.

The problem setting was described in detail in Refs. \cite{pla}, \cite{prb}.
We consider the superconductor containing columnar fragments of a normal
phase that represents either inclusions of different chemical composition or
regions with reduced superconducting order parameter. Such columnar defects
can be formed in the course of the film growth process or induced by heavy
ion bombardment \cite{mezzetti}, \cite{smith}. The columnar defects can
produce a much more intensive pinning than, for example, the point defects,
because their topology is closer to a vortex structure \cite{klaassen}, \cite
{tonomura}. The presence of columnar defects in a superconductor enhances
irreversible magnetization and, on the other hand, suppresses the magnetic
flux creep, which allows the critical current to be increased up to a value
of depairing current \cite{yeshurun}, \cite{konczykowski}.

When a sample is cooled in a magnetic field below the critical temperature,
the magnetic flux is trapped in isolated clusters of a normal phase. Then a
transport current is passed through the sample transversally to the magnetic
field. The transport current is added to all the persistent superconducting
currents, which circulate around the normal phase clusters and maintain
distribution of the trapped magnetic flux to be unchanged. By a cluster we
imply a set of columnar defects united by a common trapped flux and
surrounded by the superconducting phase. Since the distribution of the
trapped magnetic flux is two-dimensional, we will consider the transverse
cross-section of the clusters, representing extended objects, by a plane
carrying the transport current. As it was found for the first time in Ref.
\cite{pla}, the normal phase clusters can have fractal boundaries. This
feature essentially affects the dynamics of a trapped magnetic flux \cite
{prb}, \cite{tpl}, \cite{pss}. In the following, we will consider a special
case when characteristic sizes of the normal phase clusters much exceed both
the coherence length and the penetration depth. This assumption well agrees
with the data about the cluster structure of YBCO high-temperature
superconducting films \cite{pla}, \cite{prb} and is most consistent with the
important role of cluster boundaries in magnetic flux trapping.

It is assumed that a superconducting percolation cluster is formed in the
plane where the electric current flows. Such a structure provides for an
effective pinning, since the magnetic field is trapped in finite clusters of
the normal phase. As the transport current increases, a moment comes when
vortices begin to break away from the clusters of weaker pinning force than
the Lorentz force created by the current. With gradually increasing
transport current, the vortices will break away first from the clusters of a
smaller pinning force and, therefore, of a lower critical current. The
vortices will travel through the superconducting space via weak links, which
connect the clusters of the normal phase between themselves and act as the
channels for magnetic flux transport. These weak links are formed especially
readily at various structural defects in high-temperature superconductors
that have small coherence length. Various structural defects, which would
cause an additional scattering at long coherence length, give rise to the
weak links in high-temperature superconductors.

The motion of vortices broken away from pinning centers results in a voltage
drop across the sample that leads to the transition into a resistive state.
Each cluster has an individual configuration of weak links so it gives the
contribution to the total statistical distribution of critical currents. The
critical current of a cluster is proportional to the pinning force and is
equal to the current at which the magnetic flux ceases to be held inside the
normal phase cluster. Different types of the critical current distribution
for the clusters with fractal boundaries were considered in \cite{pla2},
\cite{pla1}, \cite{pss}, \cite{tpl2}.

The further consideration is restricted to the most practically important
case of an exponential-hyperbolic distribution of critical currents:
\begin{equation}
f\left( i\right) =\frac{2C}{D}i^{-2/D-1}\exp \left( -C\,i^{-2/D}\right)
\text{,}  \label{eq1}
\end{equation}
which is realized in the YBCO based film structures with exponential
distribution of the cluster areas \cite{pla}, \cite{prb}. In the expression
of Eq.~(\ref{eq1}) $i\equiv I/I_{c}$ is the dimensionless current normalized
to the critical current of the transition into the resistive state, $%
I_{c}=\alpha \left( C\overline{A}\right) ^{-D/2}$ is the transport current, $%
D$ is the fractal dimension of the cluster boundary, $C\equiv \left( \left(
2+D\right) /2\right) ^{2/D+1}$ is the constant depending on the fractal
dimension, $\overline{A}$ is the average cluster area, and $\alpha $ is the
cluster form factor.

The fractal dimension $D$ sets a scaling relationship between the perimeter $%
P$ and area $A$ of the normal phase cluster: $P^{1/D}\propto A^{1/2}$. This
relation is consistent with the generalized Euclid theorem, according to
which the ratios of corresponding geometric measures are equal when reduced
to the same dimension \cite{mandelbrot}. For the Euclidean clusters, the
fractal dimension coincides with the topological dimension of line ($D=1$),
while the dimension of fractal clusters always exceeds their topological
dimension ($D>1$) to reach maximum ($D=2$) for the clusters of the most
fractality. A fractal object has fractional dimension that reflects a highly
indented shape of its boundary \cite{mandelbrot}. Let us note that the
probability density for the exponential-hyperbolic distribution of Eq.~(\ref
{eq1}) is equal to zero at $i=0$, which implies the absence of any
contribution from negative and zero currents. That will allow us to avoid
any artificial assumption about the existence of a vortex liquid, which has
a finite resistance in the absence of transport currents because of free
vortices presence: $r\left( i\rightarrow 0\right) \neq 0$. Such an
assumption is made, for example, in the case of the normal distribution of
critical currents \cite{brown}.

The voltage drop across a superconductor in the resistive state represents
an integral response of all clusters to the action of transport current:
\begin{equation}
u=r_{f}\int\limits_{0}^{i}\left( i-i^{\prime }\right) f\left( i^{\prime
}\right) di^{\prime }  \label{eq2}
\end{equation}
where $u$ is the dimensionless voltage and $r_{f}$ is the dimensionless
resistance of the flux flow. Using the convolution integral of type (\ref
{eq2}), it is possible to find the {\it V}-{\it I} characteristics of
fractal superconducting structures for an arbitrary fractal dimension as
well as to study the dependence of resistance on the transport current. The
resistive characteristics provide important information about the nature of
the vortex state. The standard parameters in this case are the {\it dc}
(static) resistance $r\equiv u/i$, and the differential (dynamic) resistance
$r_{d}\equiv du/di$. The corresponding dimensional quantities $R$ and $R_{d}$
can be found using the formulas $R=rR_{f}/r_{f}$ and $R_{d}=r_{d}R_{f}/r_{f}$%
, where $R_{f}$ is the dimensional flux flow resistance.

For an exponential-hyperbolic distribution of critical currents described by
Equation (\ref{eq1}), expressions for the resistances of a superconductor
with fractal clusters of the normal phase are as follows:
\begin{equation}
r=r_{f}\left( \exp \left( -C\,i^{-2/D}\right) -\frac{C^{D/2}}{i}\Gamma
\left( 1-\frac{D}{2},C\,i^{-2/D}\right) \right)   \label{eq3}
\end{equation}
\begin{equation}
r_{d}=r_{f}\exp \left( -C\,i^{-2/D}\right)   \label{eq4}
\end{equation}
where $\Gamma (\nu ,z)$ is the complementary incomplete gamma-function. In
the limiting cases of the Euclidean clusters ($D=1$) and the clusters of
most fractal boundaries ($D=2$), the above formulas can be simplified:
\medskip

$D=1$:

\[
r=r_{f}\left( \exp \left( -\frac{3.375}{i^{2}}\right) -\frac{\sqrt{3.375\pi }%
}{i}%
%TCIMACRO{\func{erf}}%
%BeginExpansion
\mathop{\rm erf}%
%EndExpansion
\text{c}\left( \frac{\sqrt{3.375}}{i}\right) \right)
\]
\[
r_{d}=r_{f}\exp \left( -\frac{3.375}{i^{2}}\right)
\]
where erfc($z$) is the complementary error function, and \medskip

$D=2$:

\[
r=r_{f}\left( \exp \left( -\frac{4}{i}\right) +\frac{4}{i}%
%TCIMACRO{\func{Ei}}%
%BeginExpansion
\mathop{\rm Ei}%
%EndExpansion
\left( -\frac{4}{i}\right) \right)
\]
\[
r_{d}=r_{f}\exp \left( -\frac{4}{i}\right)
\]
where Ei($z$) is the exponential integral function.

\begin{figure}[tbp]
\epsfbox{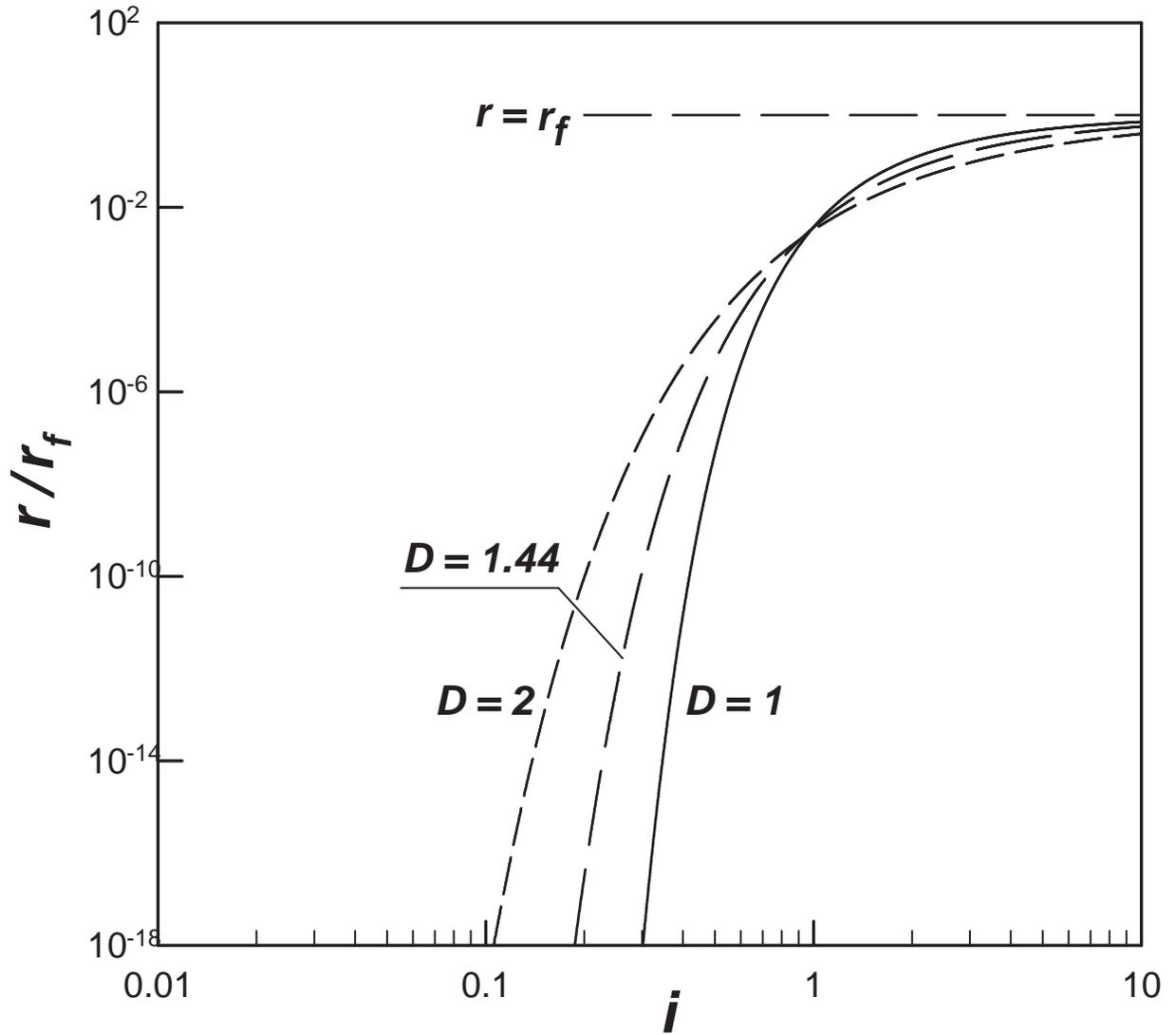}
\caption{Dependence of the {\it dc}
resistance on the transport current for superconductor with
fractal normal phase clusters ($D$ is the fractal dimension). The
dashed horizontal line $r=r_{f}$ at the upper right corresponds to
the viscous flow of a magnetic flux.} \label{figure1}
\end{figure}

\begin{figure}[tbp]
\epsfbox{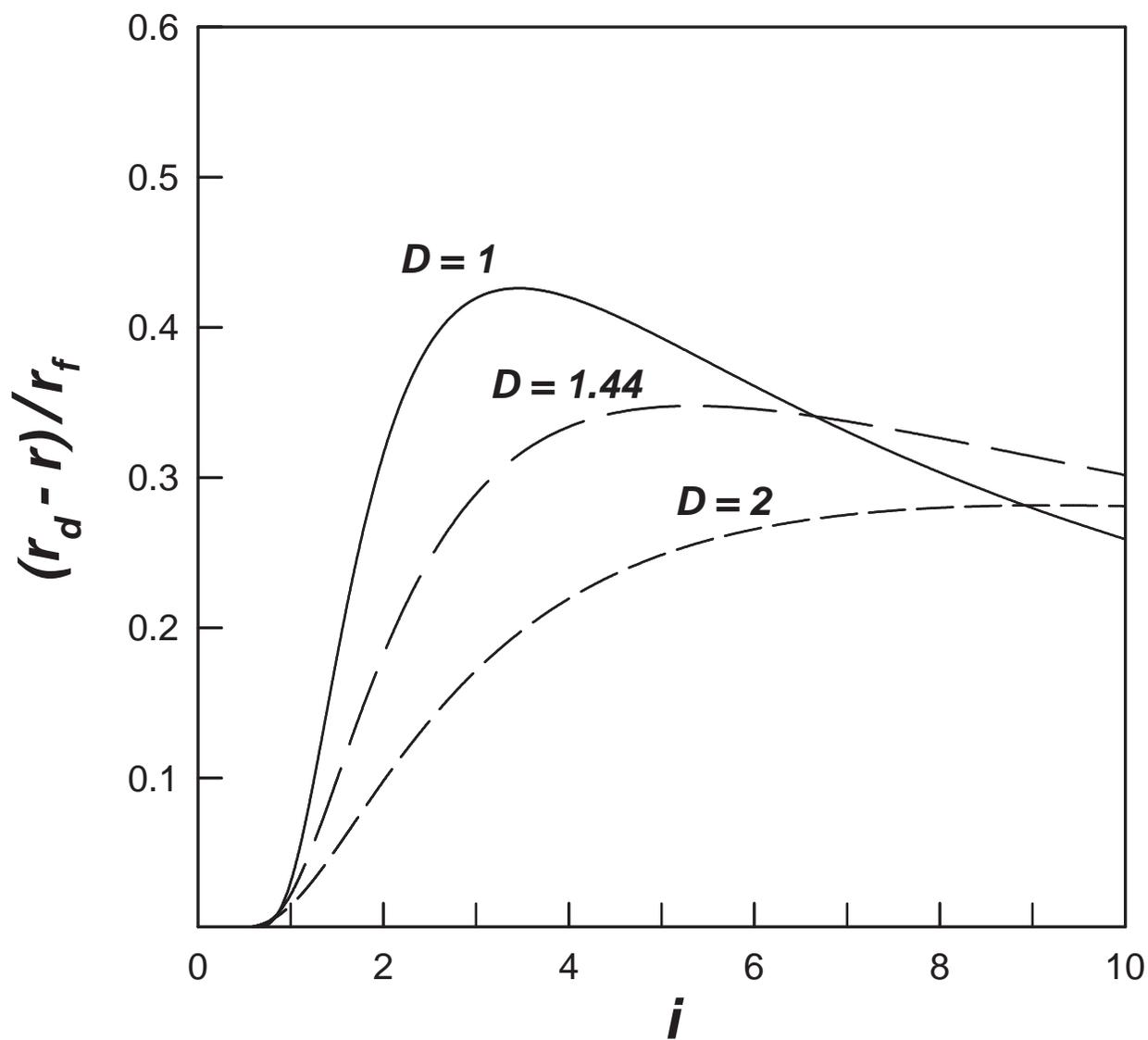}
\caption{A comparison of the {\it dc}
resistance and the differential resistance for a superconductor
with fractal normal phase clusters of various dimensions.}
\label{figure2}
\end{figure}

Figure \ref{figure1} shows the graphs of the {\it dc} resistance as a
function of transport current for superconductor with fractal clusters of a
normal phase. The curves drawn for the Euclidean clusters ($D=1$) and for
the clusters of the most fractal boundaries ($D=2$) bound a region
containing all the resistive characteristics for an arbitrary fractal
dimension. As an example, the dashed curve shows the case of the fractal
dimension $D=1.44$ found from the data of the geometric probability analysis
of the electron micrographs of YBCO superconducting film structures \cite
{pla}. The dependencies of resistance on the current shown in Fig.~\ref
{figure1} are typical of the vortex glass, whereby the curves plotted in a
double logarithmic scale are convex and the resistance tends to zero as the
transport current decreases, $r\left( i\rightarrow 0\right) \rightarrow 0$
which is related to suppression of the magnetic field creep \cite{brown},
\cite{blatter}. A vortex glass represents an ordered system of vortices,
which has no long-range ordering. At the same time, the vortex configuration
is stable in time and can be characterized by the order parameter of the
glassy state \cite{fisher}, \cite{fisher2}. In the {\it H}-{\it T} phase
diagram, mixed state of the vortex glass type exists in the region below the
irreversibility line. The dashed straight line at the upper right of Fig.~%
\ref{figure1} corresponds to a viscous flux flow regime ($r=r_{f}=const$),
which can be achieved asymptotically only. Since the {\it V}-{\it I}
characteristic of Eq.~(\ref{eq2}) of a fractal superconducting structure is
nonlinear, the {\it dc} resistance (\ref{eq3}) is not constant and depends
on the transport current. In this situation, important information can be
provided by the differential resistance, a small-signal parameter determined
by the slope of the {\it V}-{\it I} characteristic. The plots of the
differential resistance versus transport current are qualitatively analogous
to the resistance curves presented in Fig.~\ref{figure1}. A difference
between these characteristics can be seen in Fig.~\ref{figure2}. As the
fractal dimension grows, the two parameters behave more like each other, but
in the latter case there is a difference in a wider range of transport
currents. This is related to the fact that, as the fractal dimension
increases, the exponential-hyperbolic distribution of critical currents of
Eq.~(\ref{eq1}) broadens out moving toward greater i values \cite{prb}, \cite
{tpl2}.

The differential resistance is determined by the density of vortices broken
away from pinning centers by the transport current $i$,
\begin{equation}
n\left( i\right) =\frac{B}{\Phi _{0}}\int\limits_{0}^{i}f\left( i^{\prime
}\right) di^{\prime }=\frac{B}{\Phi _{0}}\exp \left( -C\,i^{-2/D}\right)
\label{eq5}
\end{equation}
where $B$ is the magnetic field, $\Phi _{0}\equiv hc/\left( 2e\right) $ is
the magnetic flux quantum, $h$ is the Planck constant, $c$ is the speed of
light, and $e$ is the electron charge. A comparison of expressions of Eqs.~(%
\ref{eq4}) and (\ref{eq5}) shows that the differential resistance is
proportional to the density of free vortices: $r_{d}=(r_{f}\Phi _{0}/B)n$.
The resistance of a superconductor in the resistive state is determined by
the motion just of these vortices because it gives rise to the voltage
across the sample.

The vortices are broken away from pinning centers mostly when $i>1$, that
is, above the resistive transition. A practically important feature of the
fractal superconducting structures consists in that the fractal character of
cluster boundaries enhances pinning \cite{tpl} and, hence, decreases the
electric field induced in the superconductor by the magnetic flux motion
\cite{pla1}. This is demonstrated both in Fig.~\ref{figure1}, where the
resistance decreases with increasing fractal dimension above the resistive
transition, and in Fig.~\ref{figure2}, where an increase in the fractal
dimension of cluster boundaries leads to a decrease in the relative
difference between the differential resistance and the {\it dc} resistance.

In the range of transport currents below the resistive transition ($i<1$),
the situation changes to opposite: the resistance increases for the clusters
of greater fractal dimension (Fig.~\ref{figure1}). This behavior is related
to the fact that, as the fractal dimension increases, the
exponential-hyperbolic distribution of critical currents of Eq.~(\ref{eq1})
broadens out covering both high and small currents. For this reason, the
breaking of the vortices away under the action of transport current begins
earlier for the clusters of greater fractal dimension. The range of currents
involved is characterized by a small number of free vortices (which is much
smaller than above the resistive transition) and, accordingly, by a low
resistance (Fig.~\ref{figure1}). This interval corresponds to the so-called
initial fractal dissipation region, which was observed in BPSCCO samples
with silver inclusions as well as in polycrystalline YBCO and GdBCO samples
\cite{prester}.

Thus, the fractal properties of the clusters of a normal phase significantly
influence the resistive transition. This phenomenon is related to the
properties of the fractal distribution of critical currents. The relations
established between the resistance and the transport current correspond to a
mixed state of the vortex glass type. An important result is that the
fractal character of the normal phase clusters enhances pinning, thus
decreasing the resistance of a superconductor above the transition into a
resistive state.

\begin{center}
${\bf Acknowledgements\smallskip }$
\end{center}

This work is supported by the Russian Foundation for Basic Researches (Grant
No 02-02-17667).

\newpage

\newpage

\end{document}